\documentclass[nonacm,sigconf]{acmart}

\usepackage{multirow}
\usepackage{hyperref}
\usepackage{xspace}

\newcommand{\os}{\textsf{OneShield}\xspace}
\newcommand{\bam}{\textsf{LLM Internal Interactive Logs}\xspace}

\setcopyright{none}

\copyrightyear{2025}
\acmYear{2025}

\begin{document}

\title{OneShield - the Next Generation of LLM Guardrails}

\author{Chad DeLuca}
\affiliation{%
  \institution{IBM Research}
  \city{San Jose}
  \country{California}}
\email{delucac@us.ibm.com}

\author {Anna Lisa Gentile}
\affiliation{%
  \institution{IBM Research}
  \city{San Jose}
  \country{California}}
\email{annalisa.gentile@ibm.com}

\author{Shubhi Asthana}
\affiliation{%
  \institution{IBM Research}
  \city{San Jose}
  \country{California}}
\email{sasthan@us.ibm.com}

\author{Bing Zhang}
\affiliation{%
  \institution{IBM Research}
  \city{San Jose}
  \country{California}}
\email{bing.zhang@ibm.com}

\author{Pawan Chowdhary}
\affiliation{%
  \institution{IBM Research}
  \city{San Jose}
  \country{California}}
\email{chowdhar@us.ibm.com}

\author{Kellen Cheng}
\affiliation{%
  \institution{Princeton University$^*$\thanks{$^*$ Work done during internship at IBM Research.}}
  \city{Princeton}
  \country{New Jersey}}
\email{kellentan@princeton.edu}

\author{Basel Shbita}
\affiliation{%
  \institution{IBM Research}
  \city{San Jose}
  \country{California}}
\email{basel@ibm.com}

\author{Pengyuan Li}
\affiliation{%
  \institution{IBM Research}
  \city{San Jose}
  \country{California}}
\email{pengyuan@ibm.com}

\author{Guang-Jie Ren}
\affiliation{%
  \institution{IBM Research}
  \city{San Jose}
  \country{California}}
\email{gren@us.ibm.com}

\author{Sandeep Gopisetty}
\affiliation{%
  \institution{IBM Research}
  \city{San Jose}
  \country{California}}
\email{sandeep.gopisetty@us.ibm.com}

\begin{abstract}

The rise of Large Language Models has created a general excitement about the great potential for a myriad of applications. While LLMs offer many possibilities, questions about safety, privacy, and ethics have emerged, and all the key actors are working to address these issues with protective measures for their own models and standalone solutions. The constantly evolving nature of LLMs makes it extremely challenging to universally shield users against their potential risks, and one-size-fits-all solutions are unfeasible. In this work, we propose OneShield, our stand-alone, model-agnostic and customizable solution to safeguard LLMs. OneShield aims to provide facilities for defining risk factors, expressing and declaring contextual safety and compliance policies, and mitigating LLM risks, with a focus on each specific customer. We describe the implementation of the framework, discuss scalability considerations, and provide usage statistics of OneShield since its initial deployment.

\end{abstract}\maketitle

%%%%%%%%%%%%%%%%%%%%%%%%%%%%%%%%%%%%%%%%%%%%%%%% INTRODUCTION %%%%%%%%%%%%%%%%%%%%%%%%%%%%%%%%%%%%%%%%%%%%%%%%
\section{Introduction}\label{sec:introduction}

%%%%%%%%%%%%%%% settings
% \alg{settings}
The advent of Large Language Models (LLMs) has opened numerous technological opportunities but also evoked apprehension from private and business users about aspects like safety, privacy, and ethics.
All major players in the field have implemented solutions to safeguard the interaction with LLMs, either as a layer to protect their own offered models or as standalone solutions to be applied to any LLM.
%%%%%%%%%%%%%%% problem
% \alg{problem}
Despite the wide variety of available guardrail solutions for LLM, and all the efforts currently going into making safety a priority\footnote{\url{https://mlcommons.org/}}, there is still a long way to go for safe AI interactions.
We identified three major gaps: (i)
many solutions are built-in with the model itself, with the risk of being self-referential and having blind spots; (ii) they are difficult to customize and/or expand towards specific enterprise needs or policy changes; (iii) the behaviors might be inconsistent, non-predictable and difficult to interpret.

%%%%%%%%%%%%%%% solution
% \alg{solution}
We propose a new generation of customizable LLM Guardrails for IBM, named \os , a collection of model-agnostic methods designed to mitigate LLM risks.
% including (i) shielding harmful or inappropriate content, (ii) ensuring the generation of accurate and reliable information within ethical considerations, (iii) mitigating biases that may be present in the training data, (iv) protecting users’ privacy, (v) identifying and correcting eventual copyright violations, (vi) verifying the factuality of generated content, and (vii) offering special handling for content about highly regulated industries such as health.
Our risk detectors fall into these three categories:
\begin{itemize}
    \item classification detectors, i.e. all those that can classify a text (typically a sentence or a paragraph) as belonging or not to a specified category. This is the case of specific domains (health, finance, etc), violence, inappropriate content, self-harm, etc.
    \item extractor detectors, i.e. those that can extract sub-portions of text in a given input that represents certain specific types of content, such as Personally Identifiable Information (PII).
    \item comparison detectors, i.e. those that can perform large-scale text retrieval and matching of specific portions of text against proprietary content - to identify data leakages or potential copyright violations, but also to verify accuracy and truthfulness of the information (factuality) against verified sources.
\end{itemize}

%%%%%%%%%%%%%%% contribution
% \alg{contribution}
The major contributions of this work are the following:
\begin{enumerate}
    \item A scalable, open, flexible, and model-agnostic architecture for a comprehensive guardrail layer for LLMs, that operates live at inference time
    \item A plethora of robust risk detectors
    % , including but not limited to: Personal Identifiable Information (PII), Hate, Abuse and Profanity (HAP), Inappropriate Content (e.g. pornography, self-harm), factuality, potential Copyright violation detection (e.g. on News, Books), leakage of proprietary information - 
    with capabilities for agile refinements and customization of each individual detector;
    \item A novel Policy Manager module that provides compliance policy templates on detection and potential violations.
\end{enumerate}

%%%%%%%%%%%%%%% advantage
% \alg{advantage}
Individually, all our risk detectors have advantages compared to currently available solutions.
For classification detectors, we design a framework that facilitates the lifetime cycle maintenance of the models and keeps them robust and up-to-date in the face of data distribution shifts. We utilize a sparse-human-in-the-loop alignment mechanism to update each detector periodically with production data.
% \cite{naacl2025STAR}. 
% Specifically - after having deployed an initial version of each detector - we run the classification task on a large enough portion of production data, and use clustering methods to speed up the human verification of the labels.
For the PII extractor, we built a novel extractor based on two primary components: an entity extractor that recognizes entities of interest and a novel Contextual PII scoring, that can identify highly sensitive data utilizing the context of each entity - we performed extensive benchmarking showing that it outperforms available state-of-the-art open-source PII detectors. 
Our comparison detectors rely on a novel \textit{text attribution} method that can compare text at scale, at any pre-defined length of words.
As a whole, the \os architecture is optimized for low-latency, inference-time guardrailing. All detectors run in parallel, guaranteeing that the total time for detection is no longer than the individual longest-running detector. Each detector can focus on a small, well-defined task exposed via a simple API, while the \os platform aggregates results for the Policy Manager to take appropriate action.
% The policy manager module interprets detectors’ annotations and provides appropriate actions based on regulations to moderate the LLM data traffic (model training, input prompts, and model responses). Customizable engines (inference/decision/action) interpret annotations and take actions (i.e. masking the text, removing it, providing user warnings, etc.) based on the currently selected policy.

% \alg{structure}
%%%%%%%%%%%%%%% Structure
The remainder of this paper is structured as follows.
In Section \ref{sec:related_work} we review the related work and present the overall framework in Section \ref{sec:method}.
Section \ref{sec:usecase} reports the implementation details for \os in several real-life use cases and end-to-end showcases, and its integration into IBM products.
% , has been used to vet large amounts of training data and discard/mark unwanted content from the training batches. Some of its components/detectors have also been integrated into IBM products - we provide details in 
% Section \ref{sec:pilot} discusses potential application scenarios and describes our pilot studies.
We discuss lessons learned and the potential evolution of this work in 
Section \ref{sec:conclusions}.

%%%%%%%%%%%%%%%%%%%%%%%%%%%%%%%%%%%%%%%%%%%%%%%% RELATED WORK %%%%%%%%%%%%%%%%%%%%%%%%%%%%%%%%%%%%%%%%%%%%%%%%
\section{Related Work}\label{sec:related_work}

Ensuring safe and trustworthy artificial intelligence 
(AI) is a mission that does not fall solely on individual technological providers, but is a mission that is recognized as a societal effort\footnote{\url{https://www.nist.gov/aisi/artificial-intelligence-safety-institute-consortium-aisic}} and the standardization of safety measures is recognized as an endeavor needed by all those involved\footnote{\url{https://mlcommons.org/benchmarks/ai-safety/general_purpose_ai_chat_benchmark/}}. 
Since the advent of LLMs, there have been efforts to categorize and define the various risks and harms posed by these technologies. Established categorizations include the taxonomy of risks by Weidinger et al. \cite{10.1145/3531146.3533088}, which clustered the risk under the macro categories that represent Discrimination, Disinformation, Misinformation, Security, Human-Computer Interaction Harms, and Environmental and Socioeconomic harms, and the categories of harm by Shelby et al. \cite{10.1145/3600211.3604673}.
Many LLMs providers recognize these potential hazards, and have accepted the accountability for ensuring responsible AI, by mitigating the various risks posed by them, and by being able to keep the interactions free from unintended responses, hallucinations, uncertainty, and compliant with ethical principles \cite{pmlr-v235-dong24c,granite2024granite,inan2023llama,rebedea2023nemo,Derczynski2024,10648691} - a very recent study proposes to create ``risk cards" as a way to standardize and democratize risk mitigation for LLMs \cite{derczynski2023assessing}.
Addressing these concerns, various key industry players have developed unique guardrail solutions, each with distinct strengths and weaknesses. 

For recent surveys on LLM safety we point the reader to \cite{dong-etal-2024-attacks,Huang2024}.
In the following we will review the most prominent mature solutions for guardrails (Section \ref{sec:industryLLM}) and the most popular datasets that are used to train and evaluate these systems (Section \ref{sec:datasets}).

\subsection{Notable LLM Guardrail Solutions}\label{sec:industryLLM}

\textbf{LlamaGuard} \cite{inan2023llama} is an open-source safeguard model tailored for Human-AI conversations, focusing on detecting unsafe prompts, mitigating prompt injections, and moderating sensitive data exchanges. Its strengths lie in adaptability and strong compliance with global privacy standards, making it well-suited for privacy-sensitive applications. However, LlamaGuard’s performance heavily depends on the quality of its training data, limiting its effectiveness in highly specialized domains. 
More importantly, LLamaGuard lacks guaranteed reliability of the results, which are tightly coupled with the model performance, since the classification of risks depends on the LLM’s understanding of the categories and the model’s predictive accuracy.
\textbf{NeMo Guardrails} \cite{rebedea2023nemo} allows developers to enforce rules that keep conversational interactions safe and on-topic. Nemo’s approach of tightly controlling interactions helps AI models avoid sensitive topics, but its reliance on predefined rules limits flexibility, requiring ongoing updates and fine-tuning. 
NeMo  utilizes similarity functions to capture the most pertinent semantics, employing the sentence transformers embeds the prompt as a vector, and then uses K-nearest neighbor (KNN) method to compare it with the stored vector-based user canonical forms.
Its effectiveness is therefore closely tied to the performance of the KNN method and to the data used as reference.
Moreover, the rule-based nature of the approach can stifle creativity and diminish the dynamic nature of AI responses if the enforcement is overly restrictive.
Google’s \textbf{ShieldGemma}~\cite{zeng2024shieldgemma} has advanced in generative AI safety by addressing harmful content, such as harassment, hate speech, and violence, through LLM-based models built on the Gemma2 framework. ShieldGemma outperforms models like LlamaGuard and WildGuard with a 10.8\% improvement on public benchmarks for harmful content detection. Its synthetic data pipeline, which augments training diversity, enables it to generalize effectively across contexts, but the rapid evolution of harmful content types poses a maintenance challenge as detection advancements must keep pace.
\textbf{Gemini Filters}\footnote{\url{https://ai.google.dev/gemini-api/docs/safety-settings}} also provide robust filtering, focusing on hate speech and misinformation with a specific emphasis on regulatory compliance. They are highly adaptable to various operational settings, though balancing flexibility with precision is challenging; overly stringent filters can reduce the fluidity of AI interactions.
 \textbf{OpenAI Moderation API}\footnote{\url{https://platform.openai.com/docs/guides/moderation/overview}} is widely used for comprehensive content moderation, scanning outputs for hate speech, harassment, and other policy violations. It employs classification techniques to flag or block outputs, offering real-time moderation. While effective in addressing overt violations, it faces challenges in nuanced contexts, leading to occasional false positives or missed subtleties in highly contextual scenarios.
\textbf{Anthropic’s Constitutional AI}\footnote{\url{https://www.anthropic.com/research/constitutional-ai-harmlessness-from-ai-feedback}} embeds ethical and safe behaviors in LLMs by guiding the model’s decisions through a set of predefined principles, or ``constitutional'' rules. These rules are designed to prevent harmful or unethical outputs while encouraging fairness, privacy protection, and transparency. Although this proactive strategy fosters ethical outputs and provides an underlying rationale, the success of this approach relies on the quality of these guiding principles. Adapting them to new requirements is resource-intensive.
\textbf{WildGuard}~\cite{han2024wildguard} leverages synthetic and real-world data to detect harmful prompts, performing well in accuracy and reliability benchmarks. 
WildGuard excels in detecting harmful content, refusal scenarios, and jailbreak attempts in LLM responses, often outperforming existing open-source tools and even competing with GPT-4 on various benchmarks. The tool's dataset is meticulously curated, enabling it to handle both direct (vanilla) and adversarial prompts with high accuracy. It achieves significant improvements in harmfulness detection, with an F1 score of 94.7\%, and in refusal detection, with an F1 score of 92.8\%, making it highly reliable for real-world applications.
However, this solution do not provide granular predictions of
harm types or only provide binary output, which limits customized harm filtering or customized thresholds for downstream use cases.
\textbf{Toolformer} \cite{schick2023toolformerlanguagemodelsteach} 
enables LLMs to automatically learn how to use external tools via API calls, and enhances LLM performance on certain tasks, such as guardrailing text, that require external knowledge or abilities not already captured within the LLM's parameters.
By relying on external safety mechanisms, it allows for a more flexible application of guardrails for different use cases, which can be decided at run-time. However the external dependencies raise some limitations, including increased system complexity, additional potential points of failure, which could bypass safeguards if the tools are misused or if the LLM mishandles these auxiliary processes.
\textbf{Guardrails AI} \cite{pmlr-v235-dong24c} encompasses three steps: defining the ``RAIL'' specification, initializing the ``guard", and wrapping the LLMs. Guardrail AI is neural-symbolic system, which consists of a backbone symbolic algorithm supported by learning algorithms (the classifier models).
It is the most similar implementation to our \os framework, where the ``RAIL'' specification is somehow equivalent to our policy manager module and the classifiers are the equivalent of our detectors. Nonetheless, \os encompasses multiple types of detectors besides the classifiers, i.e. extractors and comparisons modules.

While many recent efforts tend to incorporate the safety training within the LLM itself, it has been found that vulnerabilities persist despite the extensive red-teaming and safety-training efforts behind these models \cite{NEURIPS2023_fd661313}, and that scaling alone cannot resolve safety failure modes.
Moreover, it has been demonstrated that a malevolent user could compromise model safety via finetuning while evading detection - the Covert Malicious Finetuning method \cite{DBLP:conf/icml/Halawi0WWHS24} can infact construct a malicious dataset where every individual datapoint appears innocuous, but fine-tuning on the dataset teaches the model to respond to encoded harmful requests with encoded harmful responses.
% One way to address this is to have a ``companion" guardian model\footnote{\url{https://www.ibm.com/granite/docs/models/guardian/}} that accompanies the main LLM and addresses the safety concerns.
%
Finally, it is important to mention another important category of guardrails: the vulnerability scanners \cite{brokman2024insightscurrentgapsopensource}. A ``scanner" is a software that simulate attacks to an LLM model to evaluate its security, in the same fashion of a software penetration testing - also known as a pentest - a cyberattack on a computer system to test its security. One novel and performant scanner for LLM is garak \cite{Derczynski2024}.
While scanners are mostly an offline/one-off assessment tool,  
the guardrails suites mentioned above, act in the same fashion of a ``firewall" software in the sense that they can be used live, at inference time to intercept unwanted interactions with the LLMs.
%%%%%%%%%%%%%%%%%%%%%%%%%%%%%%%%%%%

%%%%%%%%%%%%%%%%%%%%%%%%%%%%%%%%%%%
Our \os framework is different from these efforts, being a completely independent framework, where all the detectors are small, with low computational requirements and are run in parallel, and independently from the LLM.
\os provides a cutting-edge, customizable suite for LLM guardrails, addressing a comprehensive range of AI risks. \os’s modular architecture incorporates concurrent detectors for PII, inappropriate content, factuality verification, and more, enabling real-time mitigation across multiple risk areas. With a unique Policy Manager module, \os offers regulatory templates that adapt actions such as masking or warnings based on the specific risk and applicable policies. This adaptable, scalable design makes \os suitable for various industries, especially in regulated sectors like healthcare and finance.
By integrating strengths from existing models and adding innovative features, \os distinguishes itself with its extensive compliance alignment, low-latency operation, and multimodal PII detection. The parallel detector architecture and flexible configuration make it a high-performance guardrail solution, equipping IBM with state-of-the-art protection and compliance capabilities to address the broad range of challenges associated with LLMs across domains.

%%%%%%%%%%%%%%%%%%%%%%%%%%%%%%%%%%%

%%%%%%%%%%%%%%%%%%%%%%%%%%%%%%%%%%%%%%%%%%%%%%%%%%% Experimental Setup %%%%%%%%%%%%%%%%%%%%%%%%%%%%%%%%%%%%%%%%%%%%%%%%%%

\subsection{Datasets and Benchmarks}
\label{sec:datasets}

In recent years, a wealth of data has emerged, supporting the development of robust guardrail systems for AI.
Within the \os framework, we leverage both publicly available and synthetic datasets to enhance AI safety, allowing for comprehensive training, evaluation, and testing of our detectors and framework components.
These datasets encompass a range of formats, from unstructured natural language text to semi-structured web data (e.g., tabular data), carefully curated for the distinct needs of the different modules and detectors in the framework.
We identified key datasets that provide representative samples to address our research focus.
These datasets form the foundation of our risk detectors, enabling \os to ensure LLM safety by preventing harmful, illegal, or unethical outputs and aligning synthetic data with safety standards.

\begin{itemize}
    \item AttaQ\footnote{\url{https://huggingface.co/datasets/ibm/AttaQ}}: A semi-automatically curated dataset with adversarial question attack samples. It covers harmful input prompts in seven categories, such as PII, Substance Abuse, and Violence.
    \item ALERT\footnote{\url{https://huggingface.co/datasets/Babelscape/ALERT}}: A large-scale benchmark assessing model safety across six macro and 32 micro categories, with 14,763 test prompts.
    \item SALAD-Bench\footnote{\url{https://huggingface.co/datasets/walledai/SaladBench}}: A comprehensive benchmark focusing on both attack and defense, featuring 66 categories across six domains and 16 tasks. The safety rate is calculated based on model responses classified as ``safe'' or ``unsafe.''
    \item SimpleSafetyTests\footnote{\url{https://huggingface.co/datasets/Bertievidgen/SimpleSafetyTests}}: A dataset comprising of 100 prompts across five harm areas: Suicide, Self-Harm, and Eating Disorders; Physical Harm and Violence; Illegal and Highly Regulated Items; Scams and Fraud; and Child Abuse.
    \item HarmBench\footnote{\url{https://github.com/centerforaisafety/HarmBench/tree/main}}: A standardized evaluation framework designed to test for robustness against harmful behaviors.
    \item StrongREJECT\footnote{\url{https://github.com/alexandrasouly/strongreject/tree/main}}: A collection of 313 prompts designed to evaluate the robustness of LLMs against jailbreak attempts that elicit harmful content. These prompts are categorized into six areas: Disinformation and deception, Illegal goods and services, Hate/harassment/discrimination, Non-violent crimes, Violence, and Sexual content.  
    \item AdvBench\footnote{\url{https://huggingface.co/datasets/walledai/AdvBench}}: A dataset comprising 500 adversarial prompts designed to test LLMs for vulnerabilities to harmful behaviors. These prompts cover a range of malicious activities, including instructions on creating malware, conducting cyberattacks, and inciting violence. 
    \item Do-Not-Answer\footnote{\url{https://huggingface.co/datasets/LibrAI/do-not-answer}}: A dataset comprising of 939 prompts that responsible LLMs should refrain from responding to, covering five risk areas and 12 harm types.
    \item XSTest\footnote{\url{https://huggingface.co/datasets/walledai/AdvBench}}: A dataset of 250 safe prompts that models should respond to and 200 unsafe prompts that models should appropriately refuse, facilitating the evaluation of models' calibration between helpful and safe responses.
    \item HH-RLHF\footnote{\url{https://huggingface.co/datasets/Anthropic/hh-rlhf}}: A dataset with human-generated comparisons of model outputs, focusing on helpfulness and harmlessness, and red teaming data where humans attempt to elicit harmful responses from models. 
    \item ``SuicideWatch'' Subreddit Data\footnote{\url{https://www.kaggle.com/datasets/nikhileswarkomati/suicide-watch}}:
    Human posts collected from the ``SuicideWatch'' subreddit\footnote{\url{https://www.reddit.com/r/SuicideWatch/}} on Reddit. Used for training self-harm detection models, including baseline binary and multiclass detectors.
    \item HeAL Benchmark Dataset~\cite{heal}:
    HeAL (HEalth Advice in LLMs) is a health-advice benchmark dataset that has been manually curated and annotated to evaluate LLMs' capability in recognizing health-advice to safeguard LLMs deployed in industrial settings.
    \item Enron Email Dataset~\cite{shetty2004enron}:
    Email data containing personally identifiable information (PII), such as names and phone numbers, supporting privacy guardrails.
    \item The Learning Agency Lab - PII Data Detection Dataset\footnote{\url{https://www.kaggle.com/competitions/pii-detection-removal-from-educational-data}}:
    Contains generated essays and corresponding entities used in the essay (competition entities: name, username, email, phone number, street address, url etc)
    \item LLM Generated External Dataset for PII Data Detection Dataset\footnote{\url{https://www.kaggle.com/datasets/alejopaullier/pii-external-dataset}}:
    LLM-generated (synthetic) dataset with PII information such as URLs and usernames in text data.
    \item HelpSteer (version 1 and 2)\footnote{\url{https://huggingface.co/datasets/nvidia/HelpSteer}}$^,$\footnote{\url{https://huggingface.co/datasets/nvidia/HelpSteer2}}: These datasets help align models to generate more helpful, factually correct, and coherent responses.
    \item MLCommons v0.5\footnote{\url{https://mlcommons.org/benchmarks/ai-safety/}}: Introduces an open benchmark suite for evaluating LLMs on tasks such as bias, toxicity, and factuality, enabling standardized assessment of model safety and performance.
    \item PR Insights Data:
    An internally generated benchmark from Red Hat and IBM's InstructLab platform\footnote{\url{https://github.com/instructlab}}, based on user-submitted data for risk management and guardrail insights. InstructLab~\cite{sudalairaj2024lab} is a model-agnostic open source AI project that facilitates open contributions to LLMs in an accessible way.
\end{itemize}

By integrating this diverse array of datasets, we ensure that each \os component, especially the risk detectors, is rigorously trained and tested across varied content types and contexts.
This comprehensive approach equips \os to proactively detect and mitigate risks within both input prompts and LLM-generated responses, laying a solid foundation for a robust, responsive guardrail framework.

%%%%%%%%%%%%%%%%%%%%%%%%%%%%%%%%%%%%%%%%%%%%%%%%%%% METHOD %%%%%%%%%%%%%%%%%%%%%%%%%%%%%%%%%%%%%%%%%%%%%%%%%%%
\section{\os Framework}\label{sec:method}

\begin{figure*}[t]
 \centerline{\includegraphics[width=0.85\linewidth]{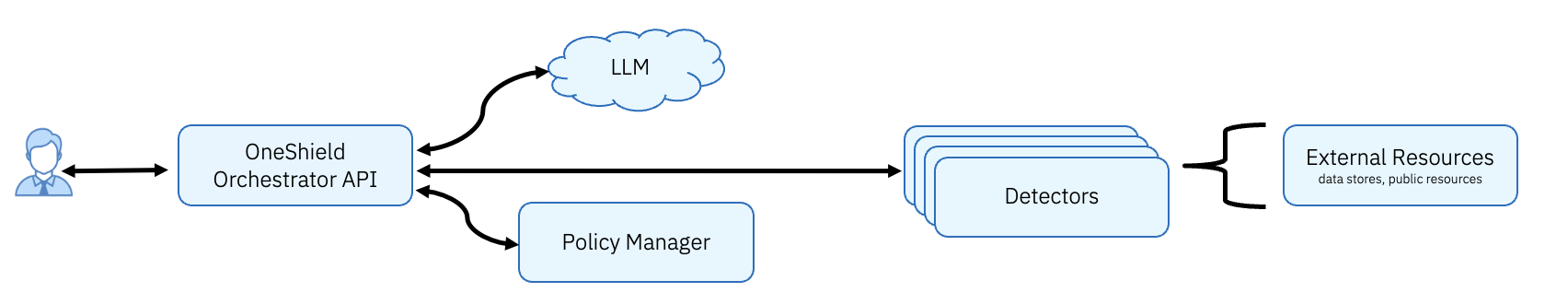}}
 \caption{OneShield Flow Diagram}
 \label{fig:flow}
\end{figure*}

The \os framework consists of a set of containerized micro-services (\ref{fig:flow} ),
% , running in Docker or Kubernetes. 
which include:
\begin{itemize}
\item \textit{\textbf{\os Orchestrator:}} The primary API and router, with endpoints for handling prompts/responses to be shielded, as well as endpoints for policy management.
\item \textbf{\textit{Detectors:}} A set of independent, stateless services, each with a specific task to classify or otherwise annotate text and return results to the orchestrator.
\item \textbf{\textit{Data Stores:}} A data store serving various corpora for retrieval/matching detectors to query directly.
\item \textbf{\textit{Policy Manager:}} A service that accepts text and aggregated detector findings to apply selected policies to the text before returning to the orchestrator.
\end{itemize}

In the following we detail the available detectors (Section \ref{sec:detectors}), the policy manager component (Section \ref{sec:policy}), and provide some general implementation details (Section \ref{sec:scale}).

\subsection{Detectors}\label{sec:detectors}
\os risk detectors fall into these three categories:
\begin{itemize}
    \item classification detectors - they label a text as belonging or not to a specified category. These detectors take as input a sentence or paragraph and output the most probable class label for it.

    \item extractor detectors - they extract sub-portions of text that represent certain specific types of content.
These detectors take as input a sentence or paragraph and output a list $P = [p_1, \dots, p_n]$ of pairs $p = (t, a)$, where $t$ is a sub-portion of the original text and $a$ is the assigned type. Currently, the set of possible types $A$ encompasses 13 Personally Identifiable Information (PII) categories.
    \item comparison detectors - they perform large-scale text comparison against proprietary content.  These detectors take as input a sentence or paragraph and output any found match against reference sources, which can be copyrighted and/or proprietary content - to identify data leakages or potential copyright violations - or openly available verified sources - to verify and prove the factuality of the information.
    
\end{itemize}
In the following section, we give an account of the detectors that have been implemented within \os.

\paragraph{Classification Detectors} 

\os classification detectors are a list of independent classifiers, currently covering the following list of risks: Health Advice.
While the \os framework allows plugging any openly available external classification model, we trained our internal models for each risk factor.
This is because the challenge in constructing effective and robust guardrails is obtaining high-quality production data. 

%%%%%%%%%%%%%%%%%%%%%%%%%%%%models

For each classifier, we trained models based on BERT embeddings \cite{devlin2018bert} constructed on
a selection of externally acquired labeled data and internal data, manually cleaned and annotated.
We utilize BERT's ability to convert input text into dense vector embeddings to capture contextual information. 
Then for the final layers, we use either (i) the \textit{BertForSequenceClassification}\footnote{\url{https://huggingface.co/docs/transformers/v4.46.2/en/model_doc/bert\#transformers.BertForSequenceClassification}} 
a pre-trained BERT \cite{devlin2018bert} model with an additional linear classification layer, or (ii) a Separable Convolutional Neural Network (SepCNN) classifier \cite{chollet2017xception}.\textit{BertForSequenceClassification} captures contextual word representations by processing entire sentences through self-attention mechanisms, whereas \textit{SepCNN} models sequences by applying depth-wise separable convolutions, focusing more on local feature extraction with fewer parameters.

%%%%%%%%%%%%%%%%%%%%%%%%%%%%data
As for the \textit{internal data} we used two sources of data: \textit{web-crawled pages}, and \bam.
\noindent For the \textit{web-crawled data}, we produce annotation with a weakly supervised process. We start from web-crawled pages and label each URL using the XForce \footnote{\url{https://www.ibm.com/x-force}}. XForce has an internal facility to classify URLs, according to a taxonomy of topics, which also includes risk categories.
We identified the classes that mapped to our target of risks in \os - and selected a predefined number of pages for each interest category.
We then use all the text content from these pages and identify salient keywords using TF-IDF analysis. We manually select the top N indicative keywords for the category and use those to identify paragraphs from the pages that explicitly contain them. This procedure produces highly focused text content for the category, making sure that we do not include boilerplate or irrelevant text in the training data. We use these paragraphs, together with the labels attached to the original URL they come from as training data points.

\noindent The \bam contains over 156 million records, including both user prompts and model responses. To select the optimal training data from the \bam, we first train a binary self-harm SepCNN classifier (baseline model) using data collected from the "SuicideWatch" subreddit on Reddit. We then apply the baseline model to the \bam, flagging data labeled as positive ("1") for further human annotation to identify false positive(FP) and true positive(TP) cases. These FPs are subsequently categorized as a third "neutral" label ("2") for use in a multiclass self-harm detection model.

% , based on real user-LLM interaction, so that the detectors work well at runtime, with real LLM generated text~\cite{ibm-detectors, wilds, gradient-detection, distribution-shift-images}. 
% \alg{\tiny The template for each detector here should be:
% \begin{itemize}
%     \item definition of the task, input and output labels
%     \item which data was used to train
%     \item results (A,P,R, F1) on which benchmark (reference or link to the benchmark is sufficient)
% \end{itemize}}

The currently available classifiers in \os are:

\begin{itemize}
    \item  \textit{Health Advice}:
    The identification of health-advice is targeted at LLM output text, where we want to interject any text that contains an explicit recommendation or suggestion on a course of actions that a person should take. 
    The detector outputs three possible labels: \textit{health-advice}, \textit{not health-advice}, and \textit{general-content}. The addition of a \textit{general-content} class helps introduce an additional layer of granularity during fine-tuning, ensuring that the predictions remain consistent for text that is not health-related.  
    We fine-tune this model using a combination of 5 academic datasets spanning both advice and health-advice recognition: NeedAdvice and AskParents~\cite{advice-on-advice}, SemEval2019-Task9~\cite{semeval2019-task9}, Detecting-Health-Advice~\cite{detecting-health-advice}, and HealthE~\cite{healthe}. As gold-standard for evaluation, we use the publicly available HeAL benchmark dataset~\cite{heal}, achieving an accuracy(A) of 85.07\%, Precision(P) of 86.64\%, Recall(R) of 88.80\% and F1 of 87.70\%.

    \item \textit{Self Harm}:
    The identification of Self Harm is targeted at LLM user prompts and model responses, where we want to detect content that promotes, encourages, or depicts acts of self-harm, such as suicide, cutting, and eating disorders. The detector outputs two possible labels: self-harm and non-self-harm. With the output labels, message the Policy Manager to take action. The training data for this self-harm detector is a balanced combination of human posts\footnote{\url{https://www.kaggle.com/datasets/nikhileswarkomati/suicide-watch}} collected from the "SuicideWatch" subreddits of the Reddit platform and the \bam. Where part of the human posts and a subset of the \bam TP are positive label "1" for self-harm-related content; and the remaining human posts are negative label "0" for non-self-harm but harmful data; a subset of the \bam FP stands for the general content with neutral label "2". Finally, the prediction is resumed into binary results for the "self-harm" category and "non-self-harm" category. This 3-class dataset is used to build a multiclass self-harm detection model, 80\% for training, 10\% for validation, and 10\% for evaluation. The evaluation results are shown in Table \ref{tab: self-harm}, details on the \textit{PR insights data} are available in Section \ref{sec:prinsights}.

    \begin{table*}
      \caption{Evaluation Results for the Self Harm Detector}
      \label{tab: self-harm}
      \begin{tabular}{p{0.12\textwidth} p{0.06\textwidth} p{0.06\textwidth} p{0.06\textwidth} p{0.06\textwidth} p{0.06\textwidth} p{0.15\textwidth} p{0.15\textwidth}}
        \toprule
        Method& Input &\multicolumn{6}{c}{SepCNN Classifier}\\
        \midrule
         Metrics & & Accuracy & Precision & Recall & F1 & False Positive Ratio & False Negative Ratio\\
        \midrule
        Test Dataset &  & 96.60\% & 96.04\% & 96.94\% & 96.49\% & 2.00\% & 3.06\%\\
        \midrule
        \multirow{3}{0.2\textwidth}{PR Insights} & Context & 99.76\% & 37.5\% & 100\% & 54.55\% & 0.24\% & 0.00\% \\
        \cmidrule(lr){2-8} & Question& 99.68\% & 29.63\% & 100\% & 45.71\% & 0.32\% & 0.00\% \\
        \cmidrule(lr){2-8} & Answer& 88.82\% & 35.29\% & 100\% & 52.18\% & 0.18\%& 0.00\% \\
        % \midrule
        % \bam & & 99.88\% & 82.18\% & 93.26\% & 87.37\% & 0.09\% &6.74\% \\
      \bottomrule
    \end{tabular}
    \end{table*}

    \item  \textit{HAP}:
This detector uses an open source model designed to detect hateful, abusive, profane, and toxic content from English text, based on training from several benchmark datasets. The model (granite-guardian-hap-125m), model card, benchmark results, and sample code can be found on HuggingFace.\footnote{\url{https://huggingface.co/ibm-granite/granite-guardian-hap-125m}}
The model is designed to operate on individual sentences, so the HAP detector chunks text into sentences and provides a confidence score for each.
        
    \item  \textit{Inappropriate Content}:
    %     \item definition of the task, input and output labels
%     \item which data was used to train
%     \item results (A,P,R, F1) on which benchmark (reference or link to the benchmark is sufficient)
% \alg{remove weapons, gambling and illegal drugs if they are not mentioned anywhere else}
    The identification of Inappropriate Content is targeted at both input and output text, where we want to interject any text that contains mentions of various category of risk, including violence, weapons, gambling, illegal drugs, sexual content amongst others.
    While we train a multi label classifier for these types of inappropriate content, we also develop binary classifiers for categories that are more prominent, e.g. sexual content.    
The training data for this detector(s) is obtained by crawling web pages, which URLs have been classified as one of the above offending categories by our IBM X-Force\footnote{\url{https://www.ibm.com/x-force}} cybersecurity product.
We then use TF-IDF analysis to identify the most salient keywords for each category and identify paragraphs in the pages containing those keywords.
For negative examples we sample content from news articles, academic and scholarly content.
For this version we focus on the adult content classifier. As gold-standard for the evaluation we use the publicly available Kaggle Adult-content-dataset\footnote{\url{https://www.kaggle.com/datasets/uyentk/valurankadult-content-dataset}}, achieving an accuracy(A) of 94.82\%, Precision(P) of 89.52\%, Recall(R) of 98.52\% and F1 of 93.80\%.
       \end{itemize}

\paragraph{Extractor Detectors}
The \os Extractor Detectors specialize in identifying and managing Personally Identifiable Information (PII). These detectors are integrated into a powerful component—available as both an integrated and standalone solution—called LLM Privacy Guard. LLM Privacy Guard leverages a multi-faceted approach, combining rule-based systems, regular expressions, and lightweight classifiers, to accurately detect and classify PII entities within text.

The tool identifies a wide range of PII types, including:
\begin{itemize}
    \item Personal details: Names, addresses, dates of birth.
    \item Contact information: Phone numbers, email addresses, social media handles.
    \item Financial data: Bank account numbers, credit card details, tax identification numbers.
    \item Government-issued identifiers: Social Security numbers (SSNs), passport numbers, driver’s license numbers.
    \item Health-related information: Medical record numbers, health insurance IDs.
\end{itemize}

Beyond detection, LLM Privacy Guard provides actionable outputs to address sensitive data, offering recommendations to mask, redact, or block PII entities. This ensures comprehensive privacy protection, aligning with global data privacy regulations and supporting secure data processing. 
% \cite{baylearn}

\paragraph{Comparison Detectors}
This category encompasses detectors that are based on matching LLMs text against proprietary content, including Knowledge Bases and document collections.
Currently, we include in this category a \textit{Text Attribution} and a \textit{Factuality} detector.
While model-based detectors rely on extensive training datasets, similarity- and search-based detectors require substantial test data as well to ensure consistent coverage and high reliability.
Through designated datasets, \os can be validated as an all-encompassing guardrail solution.

\begin{itemize}
    \item  \textit{Text Attribution}
One of the many concerns about using LLMs is the inadvertent incorporation of proprietary content that may violate copyright laws. This can happen on the model side - the generated output is based on data that has erroneously been used to train the model - as well as at the level of the prompts provided by the users - the user might unintentionally disclose proprietary data in their prompts. We implemented text-matching techniques that can vet text against proprietary content, to identify potential attribution against it. 
Our assumption is that we are in possession of the data we want to check against - this is the data collection that we want to protect, and we look for verbatim or semi-verbatim reuse of any of its content.
Scale becomes a decision factor: text similarity operations are more costly than other forms of similarity search, i.e. vector-based. 
We use vector space representation to narrow the scope of the search, and then we perform text similarity on a smaller portion of the data.
The text chunking and rewriting is performed at query time - rather than at indexing time - to allow the flexibility of searching for text of arbitrary length. The approach has been successfully used internally to vet synthetically generated data against clients' content, to avoid the unintended inclusion of confidential data in the models.
% - details of the \textit{text attribution} component can be found in \cite{naclTextAtt} - 
With respect with available state-of-the-art on LLM guardrails, this component is peculiar as it is enterprise focused and can be used to safeguard providers - not only users.

    \item  \textit{Factuality}
Factuality detection assesses whether the information in the generated text is factually correct and verifiable against external, reliable sources. It aims to detect and correct factual inaccuracies. This detector takes LLM's prompt and response as input, and executes this combined query on an external corpus(expected to contain correct answers), retrieving the most relevant passage. A BERT-based fact-checking model is developed to compare LLM's response against the retrieved results from the combined query to detect hallucinations. With the predicted binary label [yes or no], message the Policy Manager to take action. We designed 20 company intelligence-related questions, such as: "Who is Apple's current CEO?" Using these questions, we collected answers from Wikipedia for the top 100 Forbes companies. This resulted in a benchmark dataset containing 2,000 data points. This dataset serves as our evaluation benchmark. We use it to measure the model's ability to retrieve accurate, up-to-date information about companies across diverse intelligence categories. The evaluation results are accuracy(A) of 87.5\%, Precision(P) of 83.2\%, Recall(R) of 79.4\% and F1 of 81.2\%.
\end{itemize}

\subsection{Policy Manager}\label{sec:policy}

While classification and entity detection are important aspects of any guardrail system, a complete solution requires appropriate actions to be taken based on detector findings. Actions typically include allowing text to pass through unchanged, blocking text entirely, or masking certain portions while maintaining context. Decoupling the \textit{detection step} from the \textit{actions step} provides several benefits, including global scope, use-case specific actions, and jurisdictional-specific actions.

Each detector runs in parallel and is unaware of the findings of other detectors. The policy manager, which receives the aggregate of all detector findings, has access to the full context and can take actions accordingly. This allows policies to span detectors. For example, some policies may allow individual name or slightly edgy content with a moderate HAP score, but wish to block cases where an individual's name is used in a sentence with any amount of detected hate. The policy manager can easily detect this situation and block the offending text in a nuanced way, without being overly aggressive.

Policy templates provide a valuable avenue by which different use-cases and jurisdictions are accounted for, without impacting the nature of individual detectors. Regulations covering PII, hate speech, and other risky content differ regionally. For example, Europe's General Data Protection Regulation (GDPR) and the California Consumer Privacy Act (CCPA) provide specific definitions of entities and how they must be treated. These details are easily encoded in policies that can be selectively applied to ensure compliance.

\begin{figure}[]
 \centerline{\includegraphics[scale=0.47]{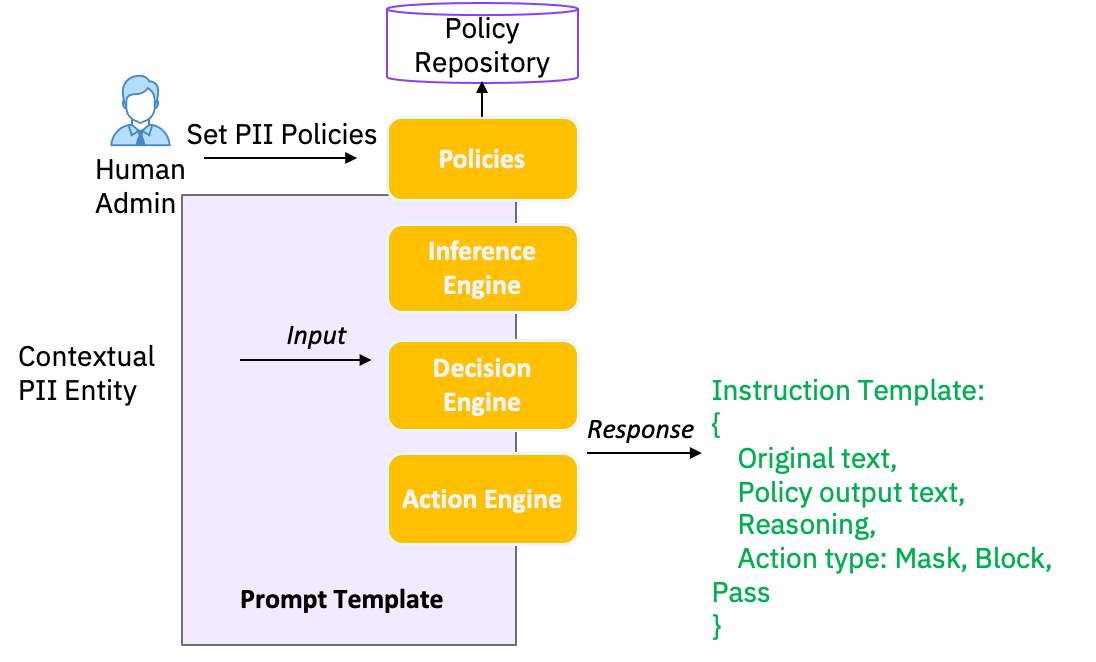}}
\caption{Policy Manager}
\label{fig:prompttemplate}
\end{figure} 

As a simple example specific to PII, Figure \ref{fig:prompttemplate} shows how various components work together to identify appropriate actions that need to be taken for a given set of entities detected according to jurisdiction. For example, the inference engine interprets the PII entities and determines the privacy level based on the policy template. The decision engine determines the action based on the entities detected and active policies. Finally, the action engine acts upon the input text to take policy defined actions.

\subsection{Scale and Implementation Considerations}\label{sec:scale}

While the services are not required to be co-located, it is beneficial for both performance and security to only expose the \os Orchestrator's endpoints, leaving the other services only accessible internally.
Scale is an important component to account for when handling run time guardrailing tasks. 
The architecture is optimized for low-latency, inference-time guardrailing. All detectors run in parallel, guaranteeing that the total time for detection is no longer than the individual longest-running detector. 
Each detector can focus on a small, well-defined task exposed via a simple API, while the \os platform aggregates results for the Policy Manager to take appropriate action.
The orchestrator and policy manager are lightweight services, each with a single instance and 40 threads. Detectors vary by type, but typically run with 5 instances and 100 threads each. As each instance must hold a model in memory (if applicable), there are performance/resource-usage trade-offs that can be made.
Every service, including those that leverage models, is optimized based on its needs and workload and supports CPU-only execution. Models will use a GPU when available, and perform significantly faster on GPU compared to CPU.
While all detectors are executed in parallel, the orchestrator waits for the final detector's response before passing results to the policy manager. This provides the policy manager with global scope and allows policies to span multiple detector findings.
As mentioned above, the total time for each call is determined by the longest running detector, which in our implementation is represented by the PII extractor, with an average response time for a user prompt of up to 150 tokens of \textit{0.521 milliseconds} (calculated on a set of 1200 user prompts), raising linearly with the number of token.
% The average response time for prompt response of up to 250 tokens was \textit{0.711 milliseconds}. 
% The guardrail also provided a template to analyze and mask the PII data as shown in Figure \ref{fig:oneshield}. This Privacy Policy Manager took an average of \textit{0.01 milliseconds} to take action like masking/blocking the input prompt or response. 

%%%%%%%%%%%%%%%%%%%%%%%%%%%%%%%%%%%%%%%%%%%%%%%% USE CASE %%%%%%%%%%%%%%%%%%%%%%%%%%%%%%%%%%%%%%%%%%%%%%%%
\section{Deployment and Release}\label{sec:usecase}
% \alg{ describing e.g., the implementation of a system, the acquisition of data, or the application of a methodology, that solves a significant real-world problem and demonstrates benefits as well as impact. Submissions must clearly describe how their work has been deployed or released, and for how long.}

\os is deployed as an integral part of our internal platform for models and data management, that provides access to all the resources surrounding AI models development.
The platform gives access to datasets, internal and external models, with up-to-date metadata, lineage, engineering insights and data governance information.
\os is deployed as a layer of risk management, both for the live interaction with all the available LLM models - highlighting risks as they arise in the user interaction - as well as a tool to inspect large quantities of data for the training and fine-tuning tasks, identifying potential vulnerabilities before the data is ingested by the models.
\os has been deployed internally as a stack in Kubernetes with an exposed REST endpoint since late 2023 and has typically handled thousands of requests daily, both from live users and programmatic requests. The deployment uses standard CI/CD tools to ensure production-ready code is pushed to the system after reviews and testing requirements are satisfied. 
Externally, the \os detectors capabilities have been utilized in the scope of an IBM open source project (Section \ref{sec:prinsights}).

\subsection{\os for InstructLab}
\label{sec:prinsights}
InstructLab\footnote{\url{https://github.com/instructlab}} is an IBM and Red Hat’s open-source project designed to lower the cost of fine-tuning large language models by allowing people to collaboratively add new knowledge and skills to any model.
InstructLab allows anyone to improve an existing LLM by fine-tuning it with additional data sources, but given the very open nature of it, we want to make sure that any data submitted by the community does not pose a risk to the model.
We used the underlying services of \os to vet the submitted data request on the public InstructLab repository.
This deployment utilized the \os guardrails solution in the GitHub repository as a safety bot. \os detectors were run against any new Pull Requests (PRs).
The PRs contain YAML files that are filled with seed examples for LLM training collected from the open-source community in the form of contexts, questions, and answers from users. Since the contributions are open, they may contain information that violates the project's Code of Conduct\footnote{\url{https://github.com/instructlab/community/blob/main/CODE_OF_CONDUCT.md}}, which defines unacceptable contributions, including but not limited to:
\begin{itemize}
    \item sexualized language or imagery
    \item trolling, insulting/derogatory comments, and personal or political attacks
    \item harassment 
    \item unathorizez sharing of private information, such as a physical or electronic address
\end{itemize}

% Since the inception of InstructLab, there have been numerous PRs coming in, which have been tracked manually by the human triage team.
A triage team of $\sim10$ people is task with ensuring that each PR does not violate the Code of Conduct - \os was leveraged to provide automatic annotation of potential violations, for the triage team to confirm. \os detectors have been deplyed in the form of a bot on the GitHub repository, that would comment and mark every potential potential violations, holding the automatic merge until human verification by the triage team. At the time of writing this paper, the \os had been run on 1200+ PRs, out of which ~8.25\% were identified as containing potential violations and confirmed by the triage team.

%%%%%%%%%%%%%%%%%%%%%%%%%%%%%%%%%%%%%%%%%%%%%%%%% CONCLUSION %%%%%%%%%%%%%%%%%%%%%%%%%%%%%%%%%%%%%%%%%%%%%%%%%
\section{Conclusions \& Future Work}\label{sec:conclusions}
The emergence of LLMs has ushered in transformative opportunities across various domains while raising critical concerns about safety, privacy, and ethical use. Addressing these challenges, we introduced \os, a model-agnostic and customizable guardrail solution designed to mitigate risks associated with LLM interactions and data management. By deploying robust risk detectors, including classification, extraction, and comparison mechanisms, \os provides comprehensive protection against harmful content, privacy violations, and factual inaccuracies.

The deployment of \os in real-world scenarios has demonstrated its scalability, effectiveness, and agility. 
Internally, it has been used to guard LLMs interaction as well as vetting training data in the internal IBM model factory.
% ,  sensitive data through its low-latency Privacy Guard, achieving superior PII detection compared to state-of-the-art solutions like Presidio Analyzer. 
Publicly, it has been successfully integrated into the InstructLab open source project, where it automated the detection of Code of Conduct violations in the community contributions, significantly reducing manual oversight, and saving valuable resources for enterprises.

The architecture of \os ensures flexibility and extensibility, enabling seamless adaptation to diverse enterprise needs and evolving policy landscapes. Its Policy Manager module empowers organizations with customizable compliance templates, supporting proactive risk management. Moreover, the solution’s ability to operate live at inference time, combined with sparse-human-in-the-loop mechanisms for continual refinement, positions \os as a pioneering framework for LLM safety and governance.

Despite its advancements, \os acknowledges the existing gaps in LLM safety, particularly the challenges of model-integrated solutions and inconsistent behaviors. Our contributions address these limitations, laying the groundwork for a new generation of transparent, customizable, and robust LLM guardrails. Moving forward, we envision expanding \os capabilities to further enhance trust and reliability in AI systems, fostering safe and ethical interactions in an increasingly AI-driven world.

\bibliographystyle{ACM-Reference-Format}
\bibliography{oneshield}

\end{document}